\documentclass[aps,a4paper,10pt,twocolumn,nofootinbib]{revtex4} 
\usepackage[T1]{fontenc}
\usepackage{mathptmx}
\usepackage{datetime}
\usepackage{amsmath}
\usepackage{amsfonts}
\usepackage{amsfonts}
\usepackage{mathrsfs}
\usepackage[mathscr]{euscript}
\usepackage[dvips]{graphicx}
\usepackage{fancyhdr}
\usepackage{colordvi}
\usepackage{hyperref} 
\usepackage{epsfig}
\usepackage{color}
\usepackage{bm}

\begin{document}

\title{A quantitative measure of carrier shocking}

\author{Paul Kinsler}
\email{Dr.Paul.Kinsler@physics.org}
\affiliation{
  Blackett Laboratory, Imperial College London,
  Prince Consort Road,
  London SW7 2AZ,
  United Kingdom.}

\begin{abstract}

I propose a definition for a ``shocking coefficient'' $S$
 intended to make determinations of the degree of waveform shocking, 
 and comparisons thereof, 
 more quantitative.
This means we can avoid having to make ad hoc judgements 
 on the basis of the visual comparison of wave profiles.

\end{abstract}




\newcommand{\sech}{{\textrm{ sech}}}

\def\FIGWIDE{0.80}

\lhead{\includegraphics[height=5mm,angle=0]{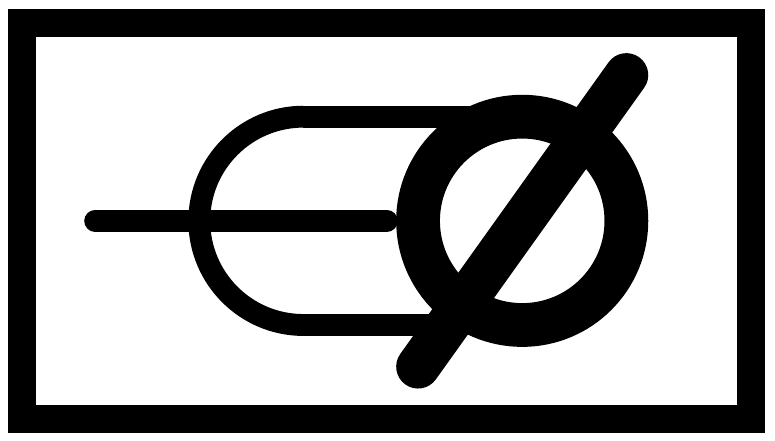}~~QSHOCK}
\chead{~}
\rhead{
\href{mailto:Dr.Paul.Kinsler@physics.org}{Dr.Paul.Kinsler@physics.org}\\
\href{http://www.kinsler.org/physics/}{http://www.kinsler.org/physics/}
}

\date{\today}
\maketitle
\thispagestyle{fancy}

\newcommand{\XDOI}[1]{\href{http://dx.doi.org/#1}{doi:#1}}
\newcommand{\XARXIV}[1]{\href{http://arxiv.org/abs/#1}{arXiv:#1}}

\chead{Shocking Coefficient}

%
\section{Introduction}
\label{S-intro}

The formation of ''shocks'' on wave profiles -- 
 regions with a step-like or
 otherwise excessively steep gradient --
 has been an interesting area of nonlinear optics 
 for some time \cite{Rosen-1965pr,DeMartini-TGK-1967pr}.
In particular, 
 recent interest has centred on carrier shocking 
 \cite{Rosen-1965pr,Flesch-PM-1996prl,Gilles-MV-1999pre,Kinsler-2007josab,Genty-KKD-2007oe,Kinsler-RTN-2007pre}
 where the underlying wave oscillations self-steepen, 
 rather than just the wave envelope \cite{DeMartini-TGK-1967pr}.
Recently it has also been shown to have potential for practical --
 or at least experimental --
 relevance, 
 based on recent predictions for and mid-infrared materials
 with realistic dispersions 
 \cite{Whalen-PKM-2014pra,Panagiotopoulos-WKM-2015arXiv}.
Consequently, 
 in this work I will consider only this
 nonlinear optical situation, 
 although the definition of shocking coefficient
 I propose may well have utility in other fields.

Here I take a purist point of view in that a shock 
 occurs (only)
 when the pulse profile has evolved to have an infinite gradient; 
 this is distinct from a more intuitive criteria involving
 the presence of a steep edge in the wave profile
 with (possibly) some fast trailing oscillations.
However, 
 by insisting on an ``infinite gradient'' criterion, 
 the notion of how ``shocked'' a given waveform or pulse
 might be is not so obvious.
After all, 
 any finite gradient, 
 however extreme, 
 is a long way from being infinitely steep.
Further, 
 even without that absolutist position, 
 how do we decide on -- 
 and most importantly \emph{quantify} --
 what features of the waveform are or might be shock-like?

One notable exception to the difficulty of discussing shocks
 occurs in numerical experiments:
 one can use the numerical difficulties that occur 
 when the pulse evolution reaches the limits of computational resolution.
This is in essence what is done  
 by the LDD (local discontinuity detection) method, 
 as already used in an optical context 
 by Kinsler et al. \cite{Kinsler-RTN-2007pre}.
However, 
 the LDD method, 
 although useful in its idealised context, 
 cannot inform us as to the degree of shocking present
 in a given waveform, 
 especially if it is \emph{not}
 approaching the limits of numerical accuracy.

In Sec. \ref{S-shockcoefficient}
 I consider a way to 
 calculate a ``shocking coefficient'' that can 
 be used as an indication as to degree of shocking present in a waveform, 
 and a as a benchmark for comparisons.
Following that, 
 in Sec. \ref{S-simulations}
 I present the results of some simulations using this coefficient, 
 and I conclude in Sec. \ref{S-conclusions}.

%
\section{Shocking coefficient}
\label{S-shockcoefficient}

Here I propose a relative measure of how shocked a given pulse is, 
 by the simple expedient of 
 regularizing the gradient with the inverse tangent function, 
 which can be used to map very large (or even infinite) gradients
 into a fixed and finite range.
I define the shocking coefficient, 
 for a point on a wave $A(t)$ of (relative) gradient $g$,
 as
~
\begin{align}
 S
&= 
  \frac{2}{\pi}
  \arctan
    \left(
      |g| - 1
    \right)
,
\label{eqn-shockcoeff}
\end{align}
 where the $2/\pi$ factor ensures that an infinitely steepening wave
 returns $S \rightarrow 1$, 
 and the $g-1$ ensures that an unchanged wave
 returns $S=0$.
If the wave has become smoother, 
 the $S$ will become negative, 
 down to a minimum of $S=-0.5$ 
 when the wave is flat, 
 i.e. $|g|=0$.
In what follows I choose the reference value
 to be the maximum value of the initial gradient, 
 as evaluated over some entire initial waveform $A_0(t)$; 
 i.e.
~
\begin{align}
  g(t)
&=
  \left[
     \frac{dA(t)}{dt}
  \right]
  \left[
    \textrm{Max}
      \left|
        \frac{dA_0(t)}{dt}
      \right|
  \right]^{-1}
.
\end{align}
Note that here I consider waves specified as functions of time $t$
 as they propagate forward in space
 \cite{Kinsler-2010pra-fchhg,Kinsler-2010pra-dblnlGpm,Kinsler-2012arXiv-fbacou}, 
 however
 there is no impediment to instead choosing to analyse wave profiles 
 defined over space using this scheme;  
 just replace $t$ above with any or all of $x, y, z$ as appropriate.
Further, 
 although there might be cases where 
 it would make sense to use a scaling that varies 
 depending on the position of a point on the wave profile, 
 I do not address that case in this work.

To demonstrate the plausibility of the definition
 in eqn. \eqref{eqn-shockcoeff},
 let us first consider a simple example.
We start with a wave of sinusoidal profile, 
 $A(t) = A_1 \sin(\omega t)$, 
 which has maximum gradient $\omega$
 and intensity $A_1^2 \omega$.
Now imagine a convenient nonlinear process
 converting this into a multifrequency wave $A'(t)$
 with a weighted spectrum of $N$ harmonics, 
 all perfectly in phase.
Note that to get every harmonic appearing
 in an optical context we would need to consider
 a suitable second order nonlinearity
 \cite{Kinsler-2007josab,Radnor-CKN-2008pra}, 
 although it is unlikely to give rise to an equal weighting 
 between the generated harmonics.
The new wave $A'$ that is to be tested for shock-like gradients
 is therefore
~
\begin{align}
  A'(t)
&=
  \sum_{m=1}^{N}
    A_m \sin(m \omega t)
.
\end{align}
Assuming an equal weighting with all $A_m=A_N$, 
 summing of the harmonics shows us that 
 this has an intensity 
~
\begin{align}
  I
&=
  \sum_{m=1}^{N}
    A_N^2 m \omega
=
  \frac{A_N^2}{2}
  N \left(N+1\right)
.
\end{align}
 which means that
 the amplitude of each harmonic --
 by conservation of energy --
 must be
~
\begin{align}
  A_N
&=
  \frac{A_1}
       {\sqrt{\frac{1}{2} N \left(N+1\right)}}.
.
\end{align}
It then follows that the maximum gradient is
~
\begin{align}
  \omega A_1 g
&=
  \sum_{m=1}^{N}
    A_N \sin \left(m \omega t\right)
=
  \frac{A_1}
       {\sqrt{\frac{1}{2} N \left(N+1\right)}}
  .
  \frac{\omega}{2}
  N \left(N+1\right)
\\
&=
  \omega {A_1}
  \sqrt{N \left(N+1\right) / 2}
.
\end{align}
Defining $N_* = \sqrt{N (N+1)/2}$, 
 and noting that for large $N$, 
 the limit is that
 $N_* \rightarrow N$, 
 the shocking coefficient for our waveform $A'$ is
~
\begin{align}
 S
&= 
  \frac{2}{\pi}
  \arctan
  \left[
      N_*
   -1
  \right]
.
\label{eqn-SofNstar}
\end{align}

As an example, 
 for waves $A'$ with $N$ up to $3, 5, 7, 9$ maximum harmonics
 (i.e. $N_*$ values of $2.45, ~3.87, ~5.29, ~6.71$)
 we then find that $S$ becomes $0.62, ~0.79, ~0.85, ~0.89$ respectively.
On the basis of this admittedly simplistic calculation, 
 we might expect than more than just the first few harmonics
 were needed if near shocking behaviour is to be indicated.
This will, 
 however, 
 depend on the threshold value of $S$ we chose to use to indicate near shocks.

Another case is when only odd harmonics are produced, 
 which (e.g.) occurs for the third-order nonlinear (Kerr) case
 typically considered in nonlinear optics and shocking
 \cite{Rosen-1965pr}.
Again making the simplifying assumption of equally weighted harmonics, 
 the same process as above can be followed.
By noting that the sum of a sequence of $M$ odd numbers is simply $M^2$,
 we find that the equivalent $N_*$ for this case is just $M$, 
 and we can reuse eqn. \eqref{eqn-SofNstar} accordingly. 
Now, 
 waves $A'$ with maximum harmonic numbers of $3, 5, 7, 9$ 
 (i.e. $N_*$ values of $2, ~3, ~4, ~5$)
 give rise to $S$ coefficients of $0.50, ~0.70, ~0.80, ~0.84$ respectively.

\begin{figure}
  \includegraphics[width=\FIGWIDE\columnwidth,angle=-0]{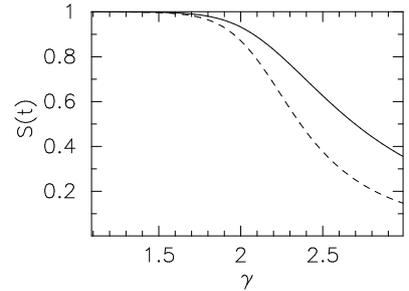}
\caption{
Shock coefficient $S$ 
 for a multi-harmonic spectrum with $A_m \propto A_1 /(m \omega_1)^\gamma$
 generated from a fundamental wave at $\omega_1$, 
 assuming energy conservation.
The solid line is the result for when all harmonics are present, 
 the dashed line is for when only odd harmonics are present. 
}
\label{fig-powlaw}
\end{figure}

Numerically we can estimate the $S$ achieved by various power law
 fall-offs in the harmonic spectrum
 under the same conservation assumption as above.
The results can be seen in fig. \ref{fig-powlaw}.
One interesting case is the $\gamma = 4/3$ fall-off considered by 
 Boyd \cite{Boyd-1992jas,Boyd-1992jcp}.
As should be expected, 
 however, 
 once the power law exponent goes beyond $-2$
 the potential for near-shocking gradients is greatly reduced.

Although these sample calculations above are indicators of how
 this shocking coefficient works, 
 the main drawback is that in realistic cases 
 the energy in the harmonics builds up gradually, 
 and we should not necessarily expect the wave energy to be redistributed
 among harmonics in some simple way; 
 not to mention the likelihood of complicated phase relationships
 being present.
Accordingly, 
 in the next section I analyse some nonlinear optical simulations 
 which provide less artificial testbed.
Nevertheless, 
 these considerations based on hamonic content
 may be useful in setting threshold values of $S$  
 that indicate nearly shocked waveforms.
For example, 
 total conversion of a wave into its third harmonic
 gives $S \simeq 0.70$ -- 
 so should we consider waves of comparable $S$ values ``shocked''?

%
\begin{figure}
  \includegraphics[width=\FIGWIDE\columnwidth]{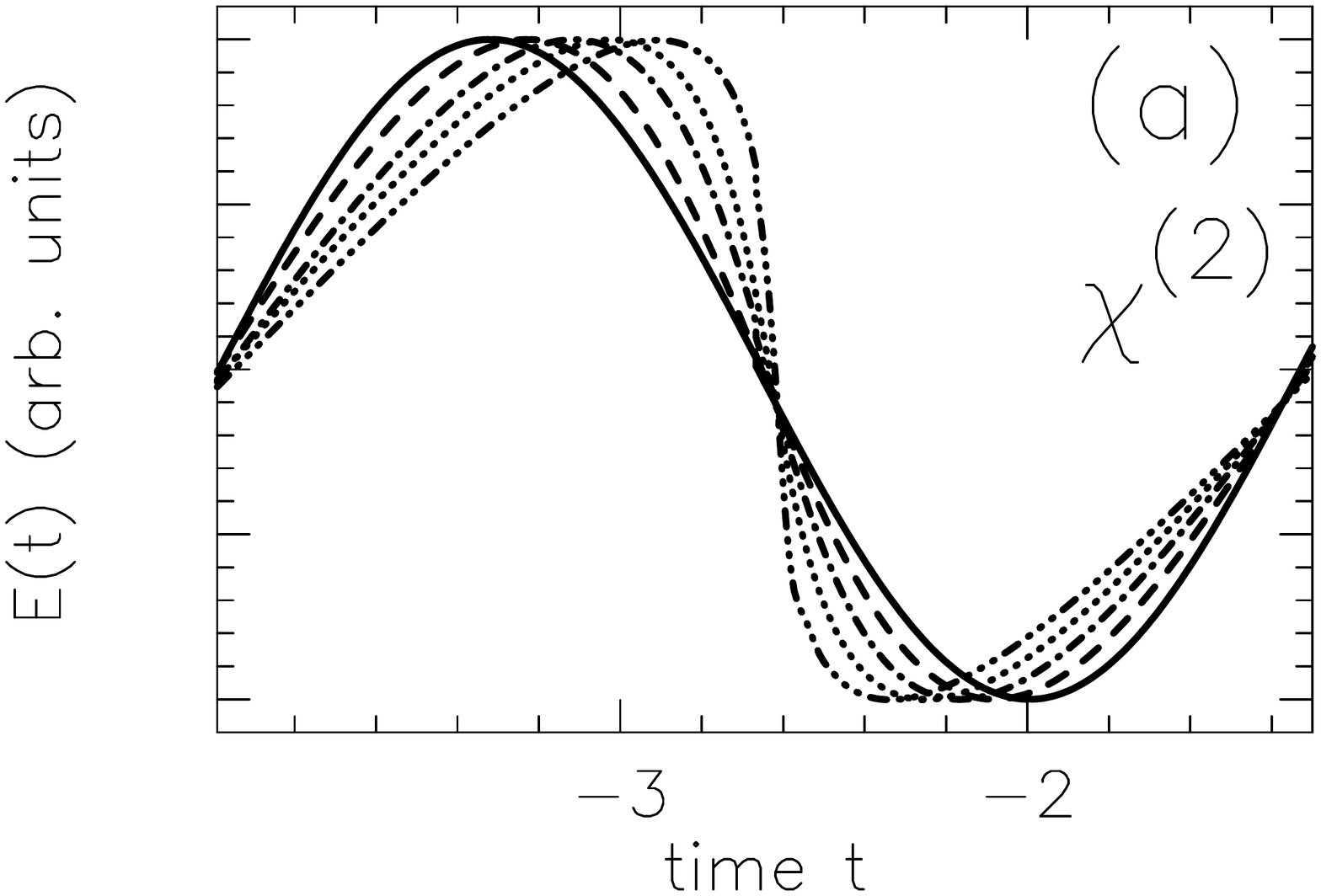}
  \includegraphics[width=\FIGWIDE\columnwidth]{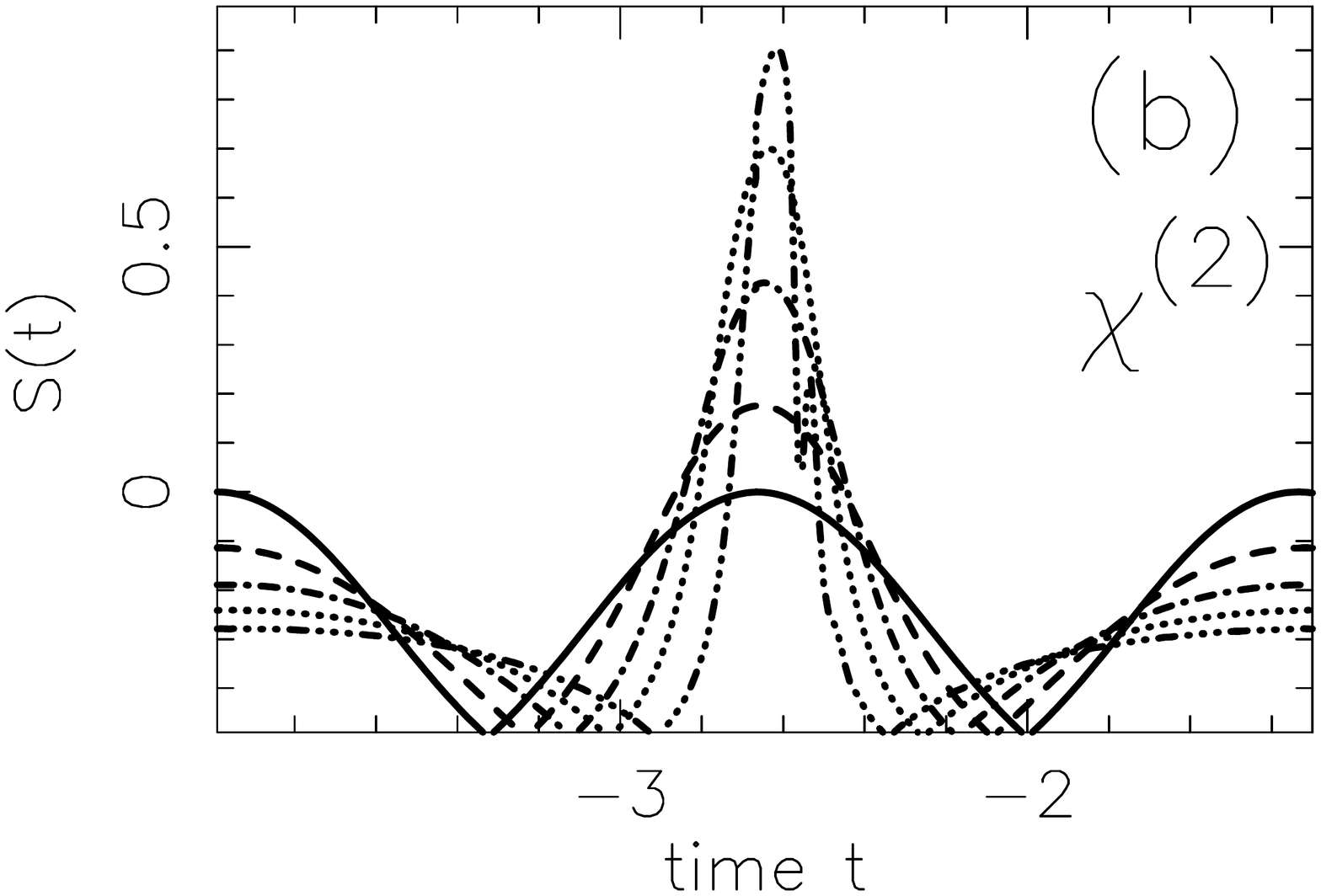}
\caption{
Shock progression in a dispersionless $\chi^{(2)}$ medium, 
 with the simulation data used to create 
 from fig.1(b) of Kinsler \cite{Kinsler-2007josab}.
(a) Wave profile, 
 where the solid, dashed, dash-dotted, and dotted lines are 
 for propagation distances closer and closer to the shocking distance.
(b) Shocking coefficient for the same data, 
 which increases as expected and also indicates the 
 location of the most shocked part of the wave. 
}
\label{fig-chi2displess}
\end{figure}

\begin{figure}
  \includegraphics[width=\FIGWIDE\columnwidth]{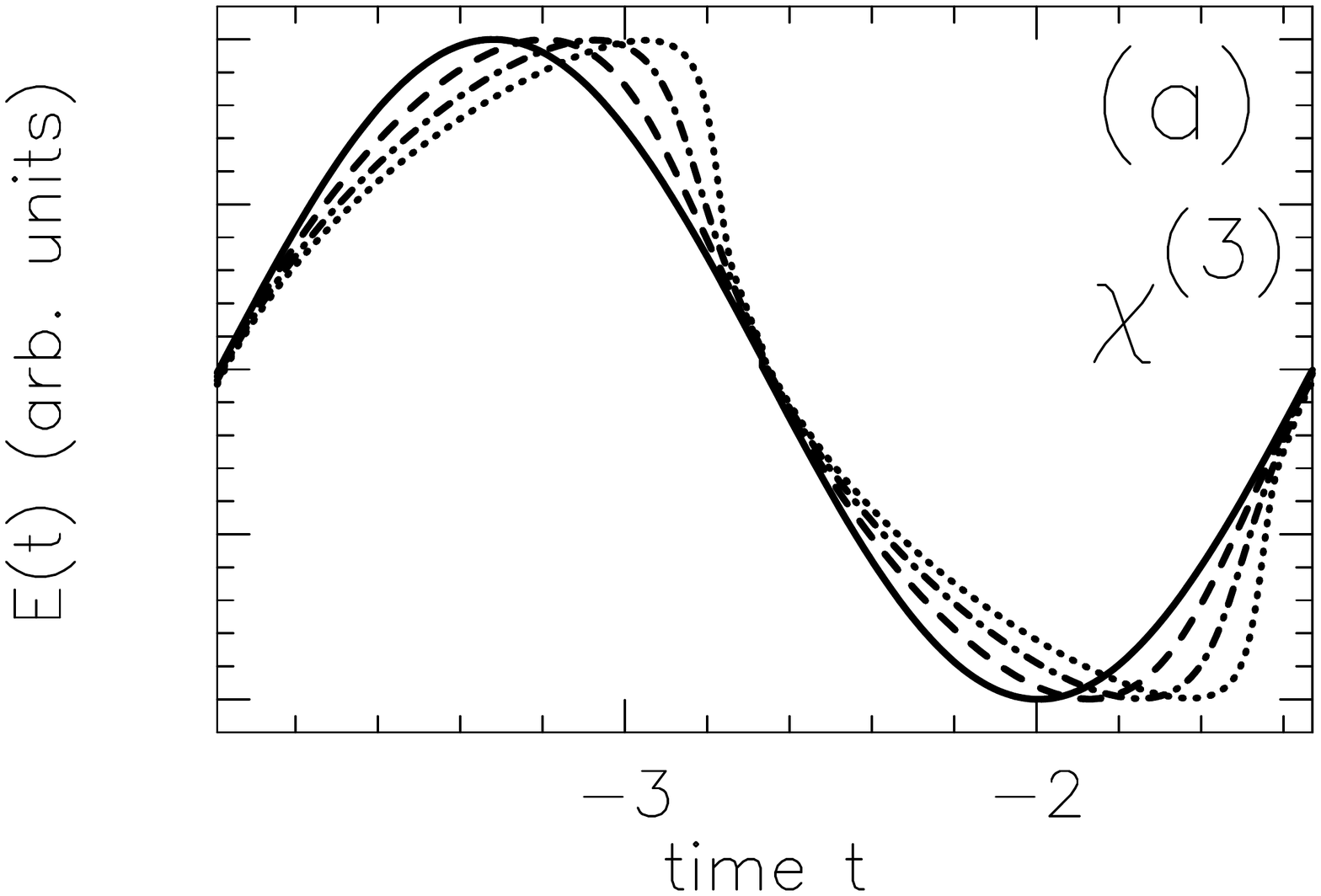}
  \includegraphics[width=\FIGWIDE\columnwidth]{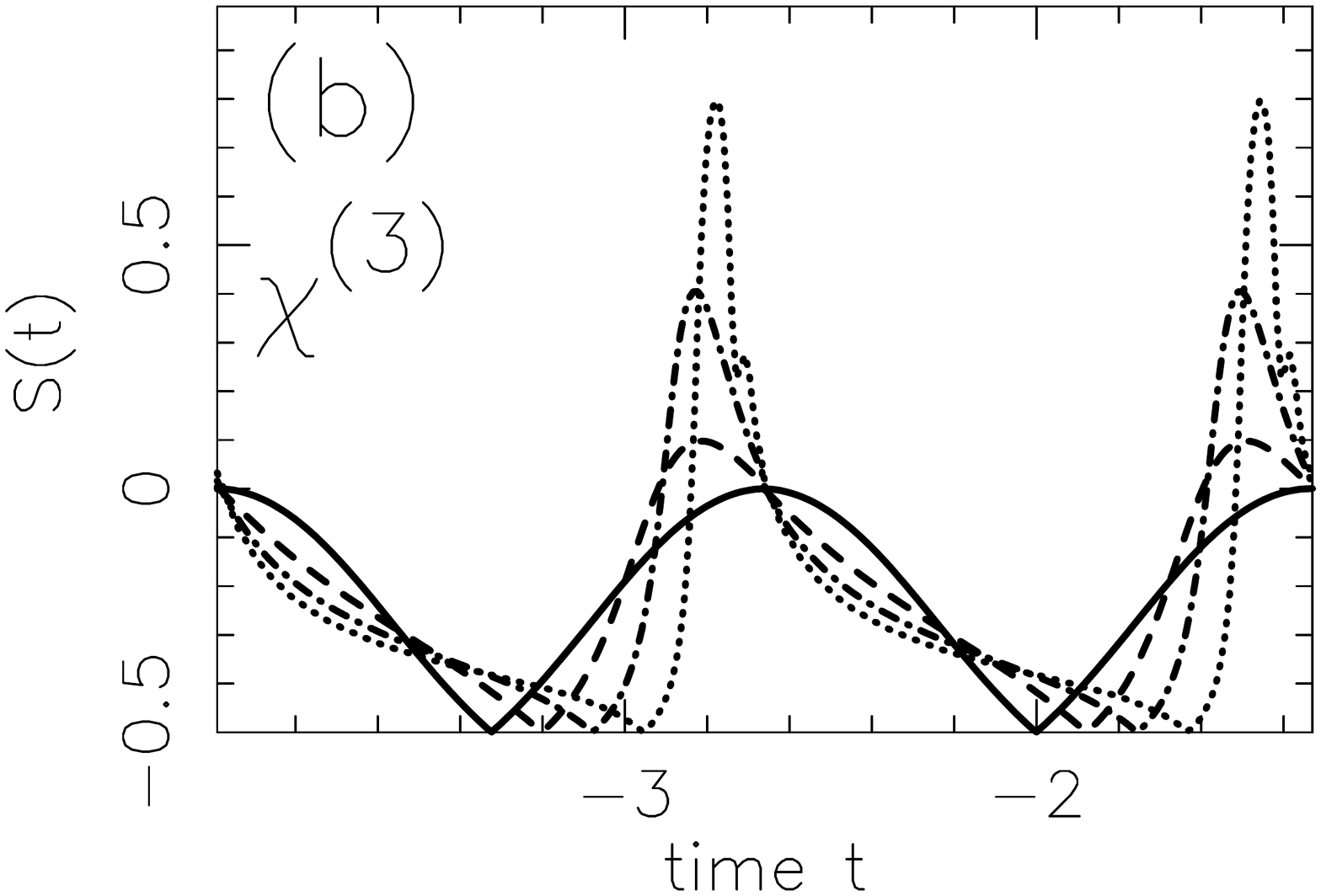}
\caption{
Shock progression in a dispersionless $\chi^{(3)}$ medium, 
 with the simulation data used to create 
 from fig.1(a) of Kinsler \cite{Kinsler-2007josab}.
(a) Wave profile, 
 where the solid, dashed, dash-dotted, and dotted lines are 
 for propagation distances closer and closer to the shocking distance.
(b) Shocking coefficient for the same data, 
 which increases as expected and also indicates the 
 moving location of the most shocked part of the wave. 
}
\label{fig-chi3displess}
\end{figure}

%
\section{Simulations}
\label{S-simulations}

The simplest simulations we might apply 
 the shocking coefficient of eqn. \eqref{eqn-shockcoeff} to
 would be for a dispersionless medium.
Accordingly, 
 in fig. \ref{fig-chi2displess}
 we show the results for such a $\chi^{(2)}$ medium, 
 based on data from Kinsler \cite{Kinsler-2007josab}
 which assumed an initial CW profile.
Fig. \ref{fig-chi2displess}(a) shows the wave profile evolution
 as the wave steepens, 
 and fig. \ref{fig-chi2displess}(b)
 show the matching increase in the peak shocking coefficient.
We also see that the increase in shocking coefficient in one
 part of the wave profile is counterbalanced by a decrease in the other part.
Further, 
 the same type of progression
 can be seen for the more commonly considered $\chi^{(3)}$ medium
 on fig. \ref{fig-chi3displess}(a,b), 
 albeit with a different wave profile evolution.

\begin{figure}
  \includegraphics[width=\FIGWIDE\columnwidth,angle=-90]{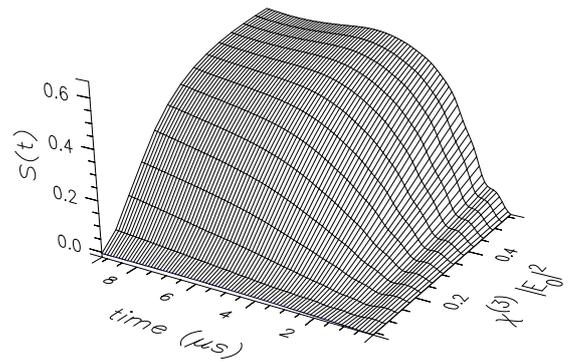}
\caption{
Shock coefficient $S(t)$ for a set of $\chi^{(3)}$ media
 with varying nonlinearity.
The basic simulation is the same
 as for fig. 1 from Genty et al. \cite{Genty-KKD-2007oe}, 
 but with fixed group velocity $v_g=1.00$ and dispersion $d_2=0.50$.
The ``missing'' $S(t)$ data at longer times and larger nonlinearities
 occurs because an LDD shock was detected and the simulation terminated.
}
\label{fig-chi3increasing}
\end{figure}

Next I turn to a less idealized situation, 
 and present the behaviour of $S$ 
 as indicated by simulations that also include dispersive effects.
Here, 
 the simulation parameters have been chosen primarily for 
 illustrative purposes.
The material response is defined by 
 a centre frequency $\omega_0 = 2.26946 \times 10^{15} s^{-1}$, 
 a refractive index $n=1.5$,
 group velocity $V_g$
 and a dispersion $D_2$
 which gives a basic quadratic form of $k(\omega)$ 
 which is :
~
\begin{align}
  k(\omega)
&=
  k_0 
 + 
  V_g 
  \left(\omega - \omega_0 \right)
 +
  \frac{1}{2}
  D_2
   \left(\omega - \omega_0 \right)^2
,
\end{align}
 although note that below I will use rescaled parameters
 $v_g = V_g \times 10^{8}$
 and 
 $d_2 = D_2 \times 10^{24}$.
However, 
 this originally specified material response undergoes
 filtering and processing to guarantee causal behaviour
 in line with the Kramers Kronig relations \cite{Kinsler-2011ejp}.

\begin{figure}
  \includegraphics[width=\FIGWIDE\columnwidth,angle=-90]{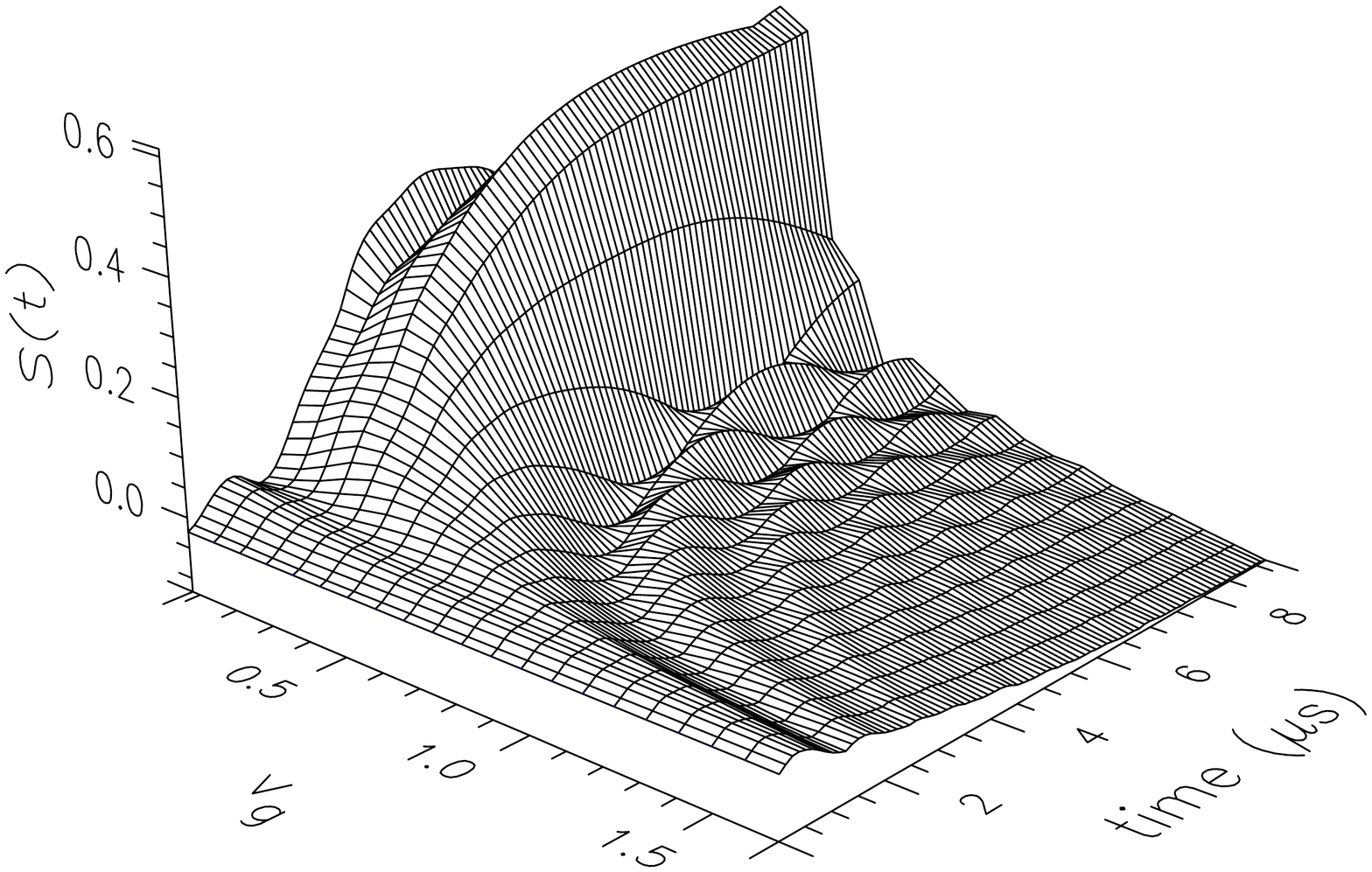}
\caption{
Shock coefficient $S(t)$ for a $\chi^{(3)}$ medium
 with a varying group velocity $v_g$.
The basic simulation is the same as for fig. \ref{fig-chi3increasing} above, 
 but with a fixed nonlinearity of $\chi^{(3)} |E_0|^2=0.32$
 and dispersion $d_2=0.50$.
}
\label{fig-chi3vg}
\end{figure}

\begin{figure}
  \includegraphics[width=\FIGWIDE\columnwidth,angle=-90]{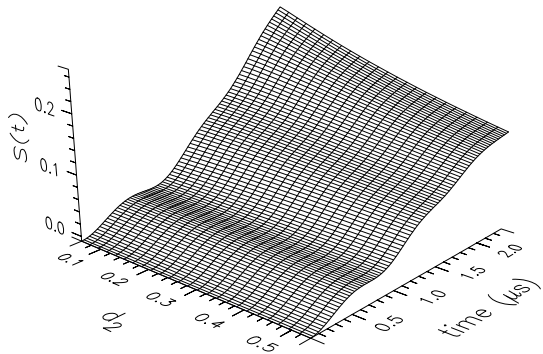}
\caption{
Shock coefficient $S(t)$ for a $\chi^{(3)}$ medium
 with a varying dispersion $d_2$.
The basic simulation is the same as for fig. \ref{fig-chi3increasing} above, 
 but with a fixed nonlinearity of $\chi^{(3)} |E_0|^2=0.32$
 and group velocity $v_g=0.60$.
For a few smaller values of $d_2$ it happens that 
 an LDD shock was detected and the simulation terminated; 
 the lack of data thereafter gives rise to the subsequent trough 
 (with an artificial $S=-0.5$)
 at larger times.
}
\label{fig-chi3dispersed}
\end{figure}

On fig. \ref{fig-chi3increasing}
 we can see how the shock coefficient $S(t)$
 changes for increasing nonlinear strength.
The behaviour is straightforward and intuitive.
For simulations where the dispersion overwhelms the nonlinearity-induced
 tendency to shock, 
 a maximum $S$ is reached after which the wave smooths out, 
 and over longer periods (than shown) $S$ is often oscilliatory in nature.
In these simulations at least, 
 when the nonlinearity is strong enough to induce a shock, 
 it happens after an interval of uniformly increasing $S$.

An alternative behaviour is shown on  fig. \ref{fig-chi3vg}, 
 where the progress of the 
 shock coefficient $S(t)$ 
 changes with an increasingly group velocity $v_g$.
We can see that once the group velocity becomes significant, 
 the maximum wave gradients as indicated by $S$ drop quickly, 
 and shocking no longer occurs; 
 the steepness also oscillates as time passes.

Lastly, 
 fig. \ref{fig-chi3dispersed}, 
 shows the progress of the 
 shock coefficient $S(t)$ 
 as it changes with an increasing dispersion $d_2$.
We can see that as is not unexpected
 \cite{Kinsler-RTN-2007pre,Panagiotopoulos-WKM-2015arXiv}, 
 the response to ever increasing dispersion 
 tends to be 
 a straightforward monotonic reduction in the maximum wave gradient.
However, 
 the interplay of nonlinearity, 
 group velocity, 
 and dispersion
 can sometimes cause shocking even when ordinarily it might not be expected.
Although not shown here, 
 some simulations with similar parameters
 to those used for fig. \ref{fig-chi3dispersed}
 did detect (LDD) shocking after short propagation distances, 
 notably for limited ranges of the smaller values of $d_2$.

The main utility of the proposed shocking coefficient $S$ 
 is that we no longer need to look as figures such as 
 fig. 1 in \cite{Radnor-CKN-2008pra} 
 or
 fig. 15 in \cite{Panagiotopoulos-WKM-2015arXiv}, 
 and \emph{by eye} try to estimate how much more or less
 a waveform is shocked compared to some comparator.
Instead we can calculate $S$ for those waves and generate results
 like those in 
 figs. \ref{fig-chi3increasing},
 \ref{fig-chi3vg},
 or \ref{fig-chi3dispersed}.
Given these, 
 it is straightforward to notice that a given change in parameters
 causes $S$ to increase or decrease by some calculable amount; 
 or that $S$ has (also) passed some well-chosen threshold value.
It may be noted that graphs showing maximum gradient of an evolving waveform
 have been presented in the past while the shocking process is being discussed
 \cite{Kinsler-RTN-2007pre,Whalen-PKM-2014pra,Panagiotopoulos-WKM-2015arXiv}, 
 but while these allow us to judge how much the field gradient has steepened, 
 they are less useful in informing us of how far they are from a \emph{shock} 
 (i.e. from infinite steepness).

As a final comment,
 all the simulations shown above were done using the computational
 model of Tyrrell at al. \cite{Tyrrell-KN-2005jmo}; 
 this is a full field approach which propagates pulses in space.
Spatial propagation methods are extremely common in nonlinear optics,
 but this is necessarily an approximate approach
 \cite{Kinsler-2010pra-fchhg,Kinsler-2012arXiv-fbacou,Kinsler-2014arXiv-negfreq,Kinsler-2015arxiv-d2owe}, 
 and caution may be needed
 in cases with extreme nonlinearity or dispersion.
However, 
 the definition of $S$ used here is not dependent
 on how a waveform was obtained, 
 and once could as easily evaluate a temporally propagated $S(x)$
 as the $S(t)$ considered here.

%
\section{Conclusions}
\label{S-conclusions}

In this paper I have proposed a quantitative measure of shocking, 
 derived from the gradient of the wave profile, 
 with the infinite steepness at shocking
 moderated down to a finite value by using 
 the inverse tangent function.
The results in sec. \ref{S-simulations} 
 suggest that that this definition of $S$
 will enable us to avoid the imprecision that follows
 a more qualitative judgement based solely on (e.g.) 
 the appearance of a wave profile.

\begin{acknowledgments}
  I acknowledge financial support from the EPSRC 
  grant number EP/K003305/1, 
 and past discussions with Mark Scaletti.
\end{acknowledgments}

%
\bibliography{/home/physics/_work/bibtex.bib}

\end{document}